# Mathematical analysis of historical income per capita distributions

Ron W Nielsen[1]

Environmental Futures Research Institute, Gold Coast Campus, Griffith University, Qld, 4222, Australia

**Abstract.** Data describing historical growth of income per capita [Gross Domestic Product per capita (GDP/cap)] for the world economic growth and for the growth in Western Europe, Eastern Europe, Asia, former USSR, Africa and Latin America are analysed. They follow closely the linearly-modulated hyperbolic distributions represented by the ratios of hyperbolic distributions obtained by fitting the GDP and population data. Results of this analysis demonstrate that income per capita was increasing monotonically. There was no stagnation and there were no transitions from stagnation to growth. The usually postulated dramatic escapes from the Malthusian trap never happened because there was no trap. Unified Growth Theory is fundamentally incorrect because its central postulates are contradicted repeatedly by data, which were used but never analysed during the formulation of this theory. The large body of readily-available data opens new avenues for the economic and demographic research. They show that certain fundamental postulates revolving around the concept of Malthusian stagnation need to be replaced by the evidence-based interpretations. Within the range of analysable data, which for the growth of population extends down to 10,000 BC, growth of human population and economic growth were hyperbolic. There was no Malthusian stagnation and there were no transitions to distinctly faster trajectories. Industrial Revolution had no impact on changing growth trajectories.

## Introduction

The aim of this publication is to present the direct proof that contrary to the fundamental postulate of the Unified Growth Theory (Galor, 2005a, 2011) distributions describing historical growth of income per capita cannot be divided into three, distinctly-different regimes of growth governed by distinctly different mechanisms. The indirect proof was presented earlier (Nielsen, 2016a, 2016c, 2016e) by showing that the historical growth of the Gross Domestic Product (GDP) and of human population were hyperbolic and that postulated by Galor three regimes of growth did not exist. Mathematical analysis of the latest data (Maddison, 2001, 2010) brings a new insight into the interpretation of the historical economic growth. Within the range of analysable data there was no Malthusian stagnation, no alleged takeoffs from stagnation to growth and no escapes from the hypothetical Malthusian trap because there was no trap.

---

Unified Growth Theory serves as a good example of traditional interpretations of economic growth revolving around the concept of Malthusian stagnation. It is also a theory, which appears to be based on Maddison's data (Maddison, 2001) but is not. Ironically, even though these excellent data were used during the formulation of this theory, they were never mathematically analysed. Unified Growth Theory is *not* based on the scientific analysis of data but on impressions supported by the habitually distorted presentation of data (Ashraf, 2009; Galor, 2005a, 2005b, 2007, 2008a, 2008b, 2008c, 2010, 2011, 2012a, 2012b, 2012c; Galor and Moav, 2002; Snowdon & Galor, 2008). Data were either used unprofessionally or they were manipulated to support preconceived ideas.

Historical economic growth and historical growth of human population were hyperbolic (Nielsen, 2016a, 2016d). Hyperbolic distributions are confusing and they are often misinterpreted in studies of economic growth and of the growth of human population. They present an image of a slow growth over a long time followed by a fast growth over a short time. These distributions are, therefore usually divided into two distinctly-different segments, slow and fast. The selected slow segment is then claimed to represent the epoch of Malthusian stagnation while the selected fast segment is assumed to represent an entirely new type of growth. The alleged transition between these two arbitrarily-selected segments is then described as explosion, takeoff, sudden spurt, sprint or the dramatic escape from the Malthusian trap. Distinctly-different mechanisms are also assigned for the two perceived segments of growth.

Often, however, interpretations of historical growth are not even based on any attempt to examine rigorously relevant data. Isolated examples are used to support the concept of stagnation followed by explosion. Even worse, more often than not, interpretations and explanations are just based on impressions and suppositions. Claims of the existence of Malthusian stagnation and transitions to different stages of growth are supported by a good dose of creative imagination.

There is no mathematically justifiable reason for dividing hyperbolic distributions into two or three distinctly-different components (Nielsen, 2014). It is mathematically *impossible* to divide hyperbolic distributions into slow and fast components. Hyperbolic distributions are slow over a long time and fast over a short time but they increase monotonically. Growth rate also increases monotonically without any unusual acceleration at any time. It increases hyperbolically with time or linearly with the size growing entity (Nielsen, 2016f). Concepts of stagnation and takeoffs from stagnation to growth are scientifically unjustifiable. They are contradicted by the analysis of data describing economic growth and the growth of population.

Hyperbolic distributions have to be interpreted as a whole and the same mechanism has to be used for the apparent slow and for the apparent fast segments. These segments do not exist even though they appear to exist. The best way to demonstrate that these segments do not exist is by using reciprocal values of hyperbolic distributions (Nielsen, 2014).

Historical economic growth is even more confusing than the historical growth of human population because economic growth is often described using income per capita represented by the Gross Domestic Product per capita (GDP/cap). It is a ratio of hyperbolic distributions and it creates an even stronger illusion of different stages of growth than the illusion created by hyperbolic distributions. It has been demonstrated (Nielsen, 2015) that the characteristic features of the GDP/cap distributions, which are interpreted as the epoch of stagnation followed by a sudden takeoff, are nothing more than mathematical properties of dividing two hyperbolic distributions. It is incorrect to claim that these features characterise uniquely economic growth.

The ratio of two hyperbolic distributions, which includes the GDP/cap ratio, increases monotonically and there is no mathematically-justifiable reason for dividing them into distinctly different regimes of growth. There is no mathematical justification for assigning different mechanisms of growth to the two perceived but non-existing segments of income per capita distributions.

Growth of income per capita was slow over a long time and fast over a short time but it was increasing monotonically. The ratio of monotonically-increasing hyperbolic distributions can only produce a monotonically-changing distribution, increasing or decreasing, depending on the singularities of hyperbolic distributions (Nielsen, 2015). Such a ratio cannot produce a distribution with a sudden discontinuity, which could be described as a takeoff.

The growth of income per capita has to be explained by using the same mechanism for the whole distribution, slow and fast. We shall now demonstrate that the empirical distributions describing income per capita were indeed increasing monotonically and that there were no sudden takeoffs from stagnation to growth as claimed incorrectly in the Unified Growth Theory (2005a, 2011).

## Unified Growth Theory

Maddison's data (Maddison, 2001, 2010) offer an unprecedented opportunity to test the past and present explanations of economic growth and of the growth of human population, explanations based on strongly-limited sources of empirical information and on creative imagination. Now, the rich body of data brings new and refreshing insights into the interpretation of the historical economic growth and of the growth of population. It is both unfortunate and ironic that Galor had access to these data but failed to use them to make important discoveries. He repeatedly distorted empirical distributions to support his preconceived ideas. An example of such distorted and self-misleading presentations of data is shown in Figure 1. However now, the same data, when properly analysed, demonstrate that the Unified Growth Theory and other similar interpretations of economic growth or the growth of population are repeatedly contradicted by empirical evidence (Nielsen, 2014, 2015, 2016a, 2016c, 2016d, 2016e, 2016f).

Hyperbolic distributions do not have to be distorted to be confusing. They are already sufficiently confusing and it is easy to make mistakes with their interpretations. Distorted presentations, such as repeatedly used by Galor, make the interpretation of these distributions even more difficult. The example presented in Figure 1 is based on a figure presented by Galor (2005a, p. 181). Such self-misleading presentations of data can be expected to lead inevitably to incorrect conclusions. It is hard to understand why such distorted diagrams were repeatedly used by Galor because the analysis of hyperbolic distributions is trivially simple (Nielsen, 2014).

The fundamental postulates of the Unified Growth Theory are based on the assumption of the existence of three, distinctly-different regimes of economic growth: Malthusian regime of stagnation, post-Malthusian regime and the sustained-growth regime. According to Galor (2005a, 2008a, 2011, 2012a), Malthusian regime of stagnation was between 100,000 BC and AD 1750 for developed regions and between 100,000 BC and AD 1900 for less-developed regions. The post-Malthusian regime was allegedly between AD 1750 and 1850 for developed regions and from 1900 for less-developed regions. The sustained-growth regime was supposed to have commenced around 1850 for developed regions.

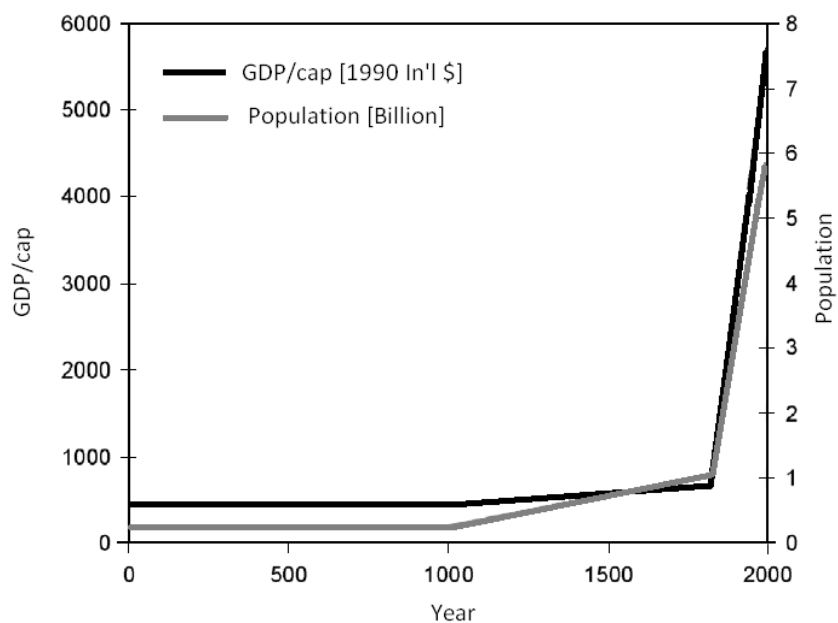

**Figure 1.** *Example of the ubiquitous, grossly-distorted and self-misleading diagrams used to create the Unified Growth Theory (Galor, 2005a, 2011). Maddison's data (Maddison, 2001) were used during the formulation of this theory but they were never analysed. Such state-of-the-art was used to construct a system of scientifically-unsupported interpretations, explanations and "mysteries of the growth process" (Galor, 2005a, p. 220).*

The end of the regime of Malthusian stagnation was supposed to have been characterised by dramatic takeoffs from stagnation to growth, described as a "remarkable" or "stunning" escape from the Malthusian trap (Galor, 2005a, pp. 177, 220). It is a signature, which cannot be missed. This change in the pattern of growth is described as "the sudden take-off from stagnation to growth" (Galor, 2005a, pp. 177, 220, 277; italics added) or as a "sudden spurt" (Galor, 2005a, 177, 220). According to Galor, the end of the Malthusian regime of stagnation for developed regions coincides with the Industrial Revolution. "The take-off of developed regions from the Malthusian Regime was associated with the Industrial Revolution" (Galor, 2005a, p. 185). Indeed, the Industrial Revolution is considered to have been "the prime engine of economic growth" (Galor, 2005a, p. 212).

For developed regions, the postulated sudden takeoffs from stagnation to growth should have occurred around AD 1750, or around the time of the Industrial Revolution, 1760-1840 (Floud & McCloskey, 1994). For less-developed regions, they should have occurred around 1900. A transition from growth to growth is not a signature of the postulated sudden takeoff from stagnation to growth. Thus, for instance, a transition from hyperbolic growth to another hyperbolic growth or to some other steadily-increasing trajectory is not a signature of the sudden takeoff from stagnation to growth. Likewise, a transition at a distinctly different time is not a confirmation of the theoretical expectations.

In the diagrams presented below, income per capita (GDP/cap) is in 1990 International Geary-Khamis dollars.

## Hyperbolic growth

It has been shown earlier that over the range of analysable data historical growth of population and historical economic growth were hyperbolic (Nielsen, 2014, 2015, 2016a, 2016b, 2016c, 2016d, 2016e, 2016f). For the economic growth, the range of analysable data extends down to AD 1 but for the growth of the world population it extends to 10,000 BC. These results are consistent with the analysis carried out over 50 years ago for the growth of the world population during the AD era (von Foerster, Mora, & Amiot, 1960) and with other similar studies (Kapitza, 2006; Kremer, 1993; Podlazov, 2002; Shklovskii, 1962, 2002; von Hoerner, 1975)

Demographic and economic research has to be based on the acceptance of hyperbolic descriptions of the historical growth of population and of the historical economic growth. Hyperbolic growth, confirmed repeatedly and consistently by data (Biraben, 1980; Clark,1968; Cook,1960; Durand, 1967, 1974, 1977; Gallant, 1990; Haub, 1995; Livi-Bacci, 1997; Maddison, 2001, 2010; McEvedy & Jones, 1978; Taeuber & Taeuber, 1949; Thomlinson, 1975; Trager, 1994) leaves no room for the outdated interpretations revolving around the concept of Malthusian stagnation followed by sudden takeoffs to a distinctly faster growth. Mathematical analysis of data consistently and repeatedly contradicts these hypothetical but unsupported concepts, including the concept that the Industrial Revolution had a decisive influence on changing growth trajectories. It did not.

Hyperbolic distribution describing growth is represented by a reciprocal of a linear function:

$$S(t) = \frac{1}{a - kt}, \qquad (1)$$

where $S(t)$ is the size of the hyperbolically growing entity (e.g. the GDP or the size of the population), while $a$ and $k$ are positive constants.

Distribution describing the time-dependence of income per capita (GDP/cap) is the ratio of two hyperbolic distributions: the hyperbolic distribution describing the growth of the GDP (Nielsen, 2016a) and the hyperbolic distribution describing the growth of population (Nielsen, 2016d). A GDP/cap distribution can be also interpreted as a ratio of two linearly decreasing distributions describing the respective reciprocal values or as a product of a hyperbolic distribution representing the GDP and a linear function representing the reciprocal values of the size of the population. Consequently, the GDP/cap ratio can be simply described as *the linearly-modulated hyperbolic distribution*, where the linear modulation is done by the reciprocal values of the size of population (Nielsen, 2015).

## Growth of the world GDP/cap

Results of mathematical analysis of the world GDP/cap are presented in Figure 2. The fitted distribution represents the linearly-modulated GDP distribution (Nielsen, 2015). Parameters describing the GDP data $a_1 = 1.684 \times 10^{-2}$ and $k_1 = 8.539 \times 10^{-6}$ while the parameters describing the world population data are $a_2 = 7.739 \times 10^0$ and $k_2 = 3.765 \times 10^{-3}$.

For the growth of the world GDP/cap we should see the signature of two takeoffs: around AD 1750 for developed regions and around AD 1900 for less-developed regions. Yet we see none of them. There was no stagnation before the Industrial Revolution and no transition from stagnation to growth around AD 1750 for developed regions or around AD 1900 for less-developed regions, as claimed by Galor (2008a, 2012a).

The data show a minor disturbance around AD 1900 which looks like a minor boosting. However, it is definitely not a "remarkable" or "stunning" escape from the Malthusian trap (Galor, 2005a, pp. 177, 220) because (1) the growth deviated only slightly from the historical trajectory, (2) this minor deviation was not preceded by stagnation and (3) because it was only temporary disturbance and the growth soon returned to the original trajectory. Furthermore, rather than being permanently and spectacularly propelled along a distinctly new trajectory, as implied by Galor's claims of "remarkable" and "stunning" takeoffs (Galor, 2005a, pp. 177, 220), economic growth as described by data, started to be diverted to a *slower* trajectory. There was definitely no transition from stagnation to growth. There was no dramatic escape from the Malthusian trap because there was no trap.

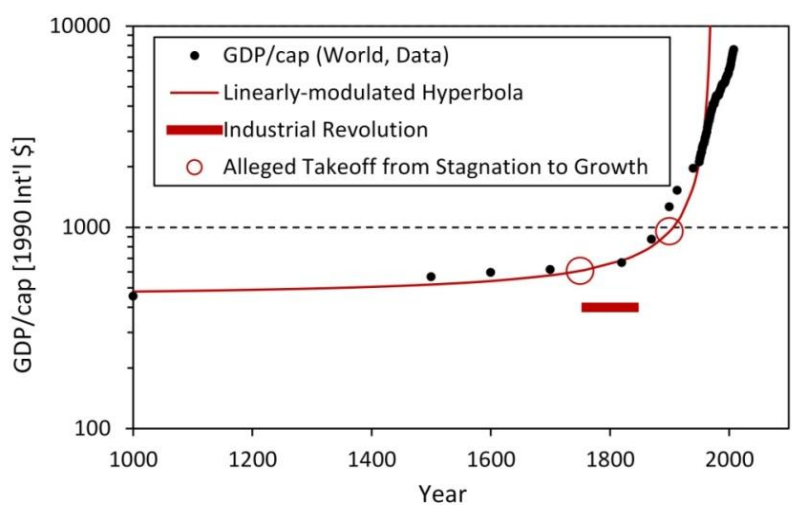

**Figure 2.** *Maddison's data (Maddison, 2010) describing the growth of the world income per capita (GDP/cap) are compared with the linearly-modulated hyperbolic distribution (Nielsen, 2015) obtained by fitting the GDP and population data (Nielsen, 2016a, 2016d). The alleged takeoffs from stagnation to growth around AD 1750 for developed regions and around 1900 for less-developed regions, as claimed by Galor (2008a, 2012a), did not happen. Industrial Revolution had no impact on changing the trajectory describing the growth of income per capita (GDP/cap).*

## Western Europe

Growth of the GDP/cap in Western Europe is shown in Figure 3. Maddison's data (Maddison, 2010) are compared with the linearly-modulated hyperbolic distribution obtained by dividing two hyperbolic distributions: the distribution describing the growth of the GDP (Nielsen, 2016a) and the distribution describing the growth of the population (Nielsen, 2016d). Parameters describing the displayed curve are: $a_1 = 9.859 \times 10^{-2}$ and $k_1 = 5.112 \times 10^{-5}$ for the GDP and $a_2 = 7.542 \times 10^1$ and $k_2 = 3.749 \times 10^{-2}$ for the growth of the population.

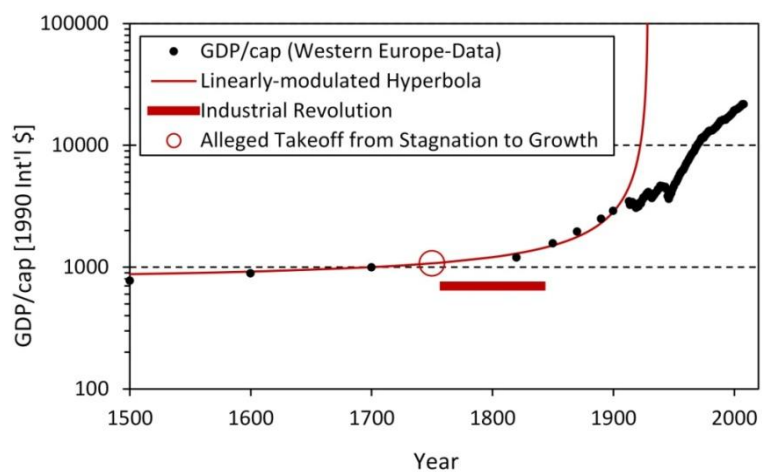

**Figure 3.** *Maddison's data (Maddison, 2010) describing the growth of income per capita (GDP/cap) in Western Europe are compared with the linearly-modulated hyperbolic distribution (Nielsen, 2015) obtained by fitting the GDP and population data (Nielsen, 2016a, 2016d). The alleged takeoff from stagnation to growth around AD 1750 (Galor, 2008a, 2012a) did not happen. Industrial Revolution had no impact on the trajectory describing the growth of income per capita (GDP/cap). The analysis of data used by Galor shows that "the prime engine of economic growth" (Galor, 2005a, p. 212) had absolutely no impact on changing the economic growth trajectory in the region where this "engine" should have been most effective and where its impacts should have been most pronounced. From around AD 1900, the growth of the GDP/cap started to be diverted to a slower trajectory.*

Results presented in Figure 3 are particularly important because they show that contrary to the generally accepted interpretations, Industrial Revolution had absolutely no impact on changing the growth trajectory of income per capita in the region where its impact should have been most pronounced. Galor's claim that the Industrial Revolution was "the prime engine of economic growth" (Galor, 2005a, p. 212) is remarkably contradicted by the same data, which he used during the formulation of his theory. This and other examples show how important Maddison's data are in correcting the outdated interpretations of the historical economic growth.

It has been shown earlier (Nielsen, 2016a) that economic growth was hyperbolic not only for the total of 30 countries of Western Europe but also for the four countries, Denmark, France, the Netherlands and Sweden, described by the most complete sets of data and representing the most advanced economies. For these countries, hyperbolic growth was between AD 1 and 1875 when it started to be diverted to a *slower* trajectory. There was no Malthusian stagnation, no takeoff and no escape from the Malthusian trap, because there was no trap. Industrial Revolution had absolutely no impact on changing economic growth trajectory in these four progressive countries where the impact of this revolution should be clearly demonstrated in the economic growth data.

Analysis of Maddison's data (Maddison, 2010) demonstrates that the "remarkable" or "stunning" escape from the Malthusian trap (Galor, 2005a, pp. 177, 220) never happened because there was no trap. Whether expressed in terms of the GDP or GDP/cap, economic growth was remarkably undisturbed during the time of the Industrial Revolution and continued undisturbed until around 1900, when it started to be diverted to a slower trajectory.

## Eastern Europe

Results of analysis of the growth of income per capita in Eastern Europe are summarized in Figure 4. Maddison's data (Maddison, 2010) are compared with the linearly-modulated hyperbolic distribution obtained by dividing two hyperbolic distributions: the distribution describing the growth of the GDP (Nielsen, 2016a) and the distribution describing the growth of population (Nielsen, 2016d). Parameters describing the fitted GDP/cap distribution are: $a_1 = 7.749 \times 10^{-1}$ and $k_1 = 4.048 \times 10^{-4}$ for the GDP and $a_2 = 3.055 \times 10^2$ and $k_2 = 1.525 \times 10^{-1}$ for the growth of the population.

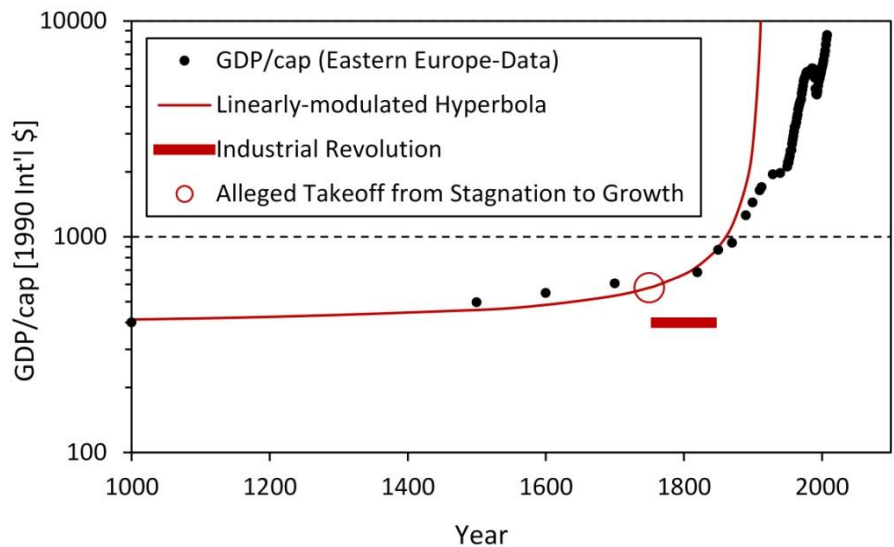

**Figure 4.** *Maddison's data (Maddison, 2010) describing the growth of income per capita (GDP/cap) in Eastern Europe are compared with the linearly-modulated hyperbolic distribution (Nielsen, 2015) obtained by fitting the GDP and population data (Nielsen, 2016a, 2016d). The alleged takeoff from stagnation to growth around AD 1750 (Galor, 2008a, 2012a) did not happen. Industrial Revolution had no impact on the trajectory describing the growth of income per capita (GDP/cap). From around AD 1850, rather than being boosted by the Industrial Revolution, the growth of the GDP/cap started to be diverted to a slower trajectory.*

Growth of income per capita was slow but it was not stagnant. It was following the linearly-modulated hyperbolic distribution. Industrial Revolution had absolutely no impact on shaping the growth trajectory. The "stunning" takeoff postulated by Galor did not happen. His theory is repeatedly and consistently contradicted by the data he used during the formulation of his theory. Rather than being boosted by the Industrial Revolution, the growth of the GDP/cap started to be diverted to a *slower* trajectory from as early as around AD 1850.

## Former USSR

Results of analysis of the growth of income per capita in the former USSR are summarized in Figure 5. Maddison's data (Maddison, 2010) are compared with the linearly-modulated hyperbolic distribution obtained by dividing two hyperbolic distributions: the distribution describing the growth of the GDP (Nielsen, 2016a) and the distribution describing the growth of population (Nielsen, 2016d). Parameters describing the fitted GDP/cap distribution are: $a_1 = 6.547 \times 10^{-1}$ and $k_1 = 3.452 \times 10^{-4}$ for the GDP and $a_2 = 2.618 \times 10^2$ and $k_2 = 1.333 \times 10^{-1}$ for the growth of the population.

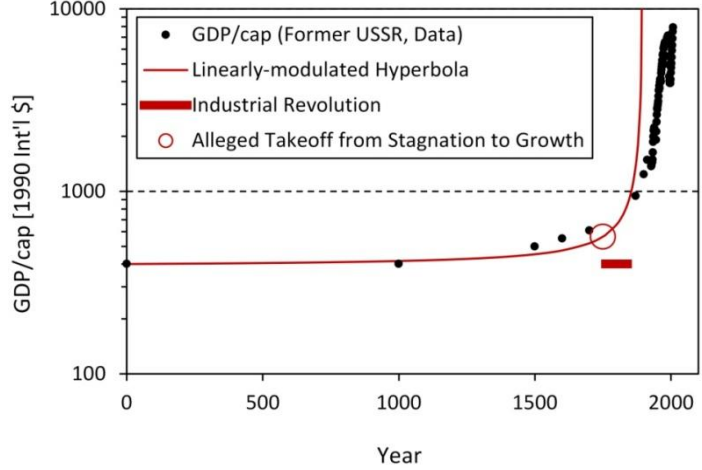

**Figure 5.** *Maddison's data (Maddison, 2010) describing the growth of income per capita (GDP/cap) in the former USSR are compared with the linearly-modulated hyperbolic distribution (Nielsen, 2015) obtained by fitting the GDP and population data (Nielsen, 2016a, 2016d). The alleged takeoff from stagnation to growth around AD 1750 (Galor, 2008a, 2012a) did not happen. Industrial Revolution had no impact on the trajectory describing the growth of income per capita (GDP/cap). From around AD 1870, rather than being boosted by the Industrial Revolution, the growth of the GDP/cap started to be diverted to a slower trajectory.*

Growth of income per capita in the countries of the former USSR was following closely the linearly-modulated hyperbolic distribution from AD 1. The growth was slow but not stagnant. Growth of the GDP and population were monotonic (Nielsen, 2016a, 2016d) and consequently the growth of income per capita (GDP/cap) was also monotonic. The "remarkable" or "stunning" takeoff (Galor, 2005a, pp. 177, 220) claimed by Galor never happened. This wished-for feature is repeatedly contradicted by the analysis of economic and population data (Nielsen, 2016a, 2016b, 2016c, 2016d, 2016e, 2016f) and by the analysis of the GDP/cap distributions. Soon after the alleged, but non-existent sudden takeoff from the non-existent stagnation to growth, the growth of income per capita in the countries of the former USSR started to be diverted to a new and *slower* trajectory.

## Asia

Analysis of the growth of income per capita (GDP/cap) in Asia (including Japan) is summarised in Figure 6. Maddison's data (Maddison, 2010) are compared with the linearly-modulated hyperbolic distribution obtained by dividing two hyperbolic distributions: the distribution describing the growth of the GDP (Nielsen, 2016a) and the distribution describing the growth of the population (Nielsen, 2016d). Parameters describing the fitted GDP/cap distribution are: $a_1 = 2.303 \times 10^{-2}$ and $k_1 = 1.129 \times 10^{-5}$ for the GDP and $a_2 = 1.068 \times 10^1$ and $k_2 = 4.999 \times 10^{-3}$ for the growth of population.

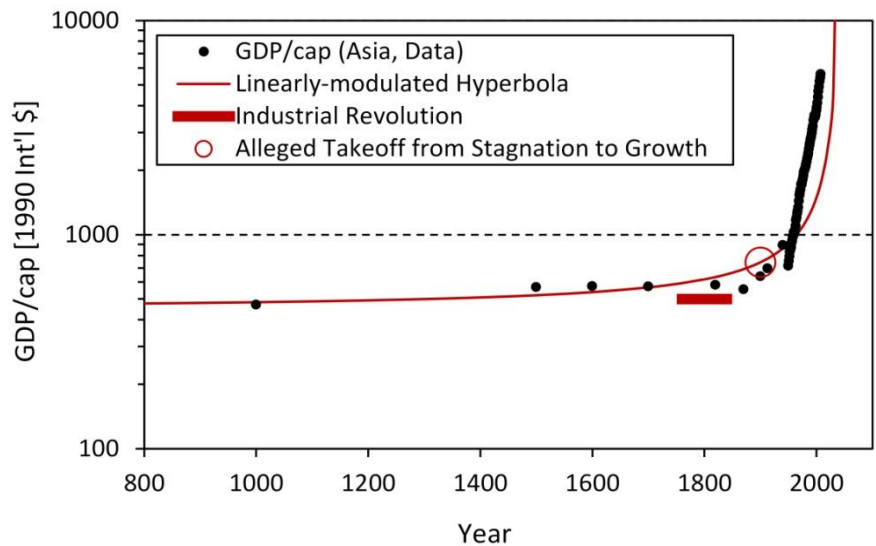

**Figure 6.** *Maddison's data (Maddison, 2010) describing the growth of income per capita (GDP/cap) in Asia are compared with the linearly-modulated hyperbolic distribution (Nielsen, 2015) obtained by fitting the GDP and population data (Nielsen, 2016a, 2016d). The boosting, which commenced around AD 1950 follows closely the original historical trajectory and is likely to cross it and move to the other side. This temporary disturbance is a part of the commonly-observed recent transitions to slower trajectories. The only difference in this case is that the transition to a slower trajectory was preceded by a minor boosting.*

Growth of the GDP/cap was slow over a long time, which is hardly surprising because the initial slow growth is the mathematically-characteristic feature of the GDP/cap distributions (Nielsen, 2015). The growth was following closely the linearly-modulated hyperbolic trajectory determined by dividing the hyperbolic distribution fitting the GDP data (Nielsen, 2016a) by the hyperbolic distribution fitting the population data (Nielsen, 2016d).

Asia is made primarily of less-developed countries (BBC, 2014; Pereira, 2011) so the alleged "stunning" takeoff from the alleged stagnation to growth should have occurred around AD 1900 (Galor, 2008a, 2012a). The data show a certain degree of boosting shortly after the time of the claimed "stunning" takeoff from stagnation to growth. However, this boosting is not a transition from stagnation to growth because the preceding trajectory was not stagnant and because the boosted trajectory follows closely the historical trend. It was obviously only a temporary boosting because the boosted trajectory is progressively coming closer to the historical trajectory and judging by its decreasing gradient it is likely to move to the other side.

This boosting could be probably explained by Japan's contribution to the total GDP/cap. Until 1900, Japan's contribution was less than 5% but by 1950 it gradually increased to 12% and by 2000 it climbed to 20%. Japan belongs to the more-developed countries so according to Galor (2008a, 2012a) it should have experienced "remarkable" and "stunning" takeoff (Galor, 2005a, pp. 177, 220) in its GDP/cap around 1750 but it did not. On the other hand, Asia should have experienced a sudden explosion in the GDP/cap growth around 1900 but it did not. There was no dramatic transition from stagnation to growth as claimed by Galor but only a transition from the non-stagnant, linearly-modulated hyperbolic trajectory to a temporarily faster growth, which appears to have been caused primarily, if not entirely, by the increasing contribution of Japan's economy, the contribution, which should have commenced explosively around 1750 but it did not. Impressions prompted by wished-for features and reinforced by distorted presentations of data such as shown in Figure 1 could be persuasive but they can be also strongly misleading. Data have to be rigorously analysed.

## Africa

Results of analysis of the growth of income per capita in Africa are presented in Figure 7. As demonstrated earlier (Nielsen, 2016a, 2016d), the GDP and population data for Africa can be fitted using two hyperbolic distributions, a slow distribution followed by a fast distribution. The transition from the slow to fast distribution occurred around 1820 for the growth of the GDP and around 1840 for the growth of population.

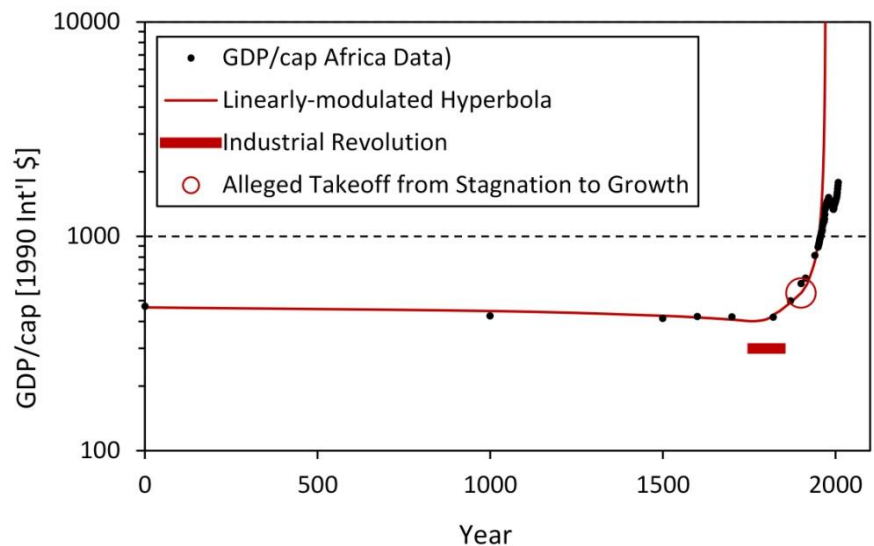

**Figure 7.** *Maddison's data (Maddison, 2010) describing the growth of income per capita (GDP/cap) in Africa are compared with the linearly-modulated hyperbolic distributions (Nielsen, 2015) obtained by fitting the GDP and population data (Nielsen, 2016a, 2016d). The alleged takeoff from stagnation to growth around AD 1900 for less-developed regions (Galor, 2008a, 2012a) did not happen because the GDP/cap trajectory was not stagnant before that year and because it continued undisturbed after this year until around 1950 when it started to be diverted to a slower trajectory.*

Parameters describing the fitted GDP/cap distribution between AD 1 and 1820 are: $a_1 = 1.244 \times 10^{-1}$ and $k_1 = 5.030 \times 10^{-5}$ for the GDP and $a_2 = 5.794 \times 10^1$ and $k_2 = 2.473 \times 10^{-2}$ for the growth of the population. For the GDP/cap distribution from AD 1840, parameters are: $a_1 = 4.192 \times 10^{-1}$ and $k_1 = 2.126 \times 10^{-4}$ for the GDP and $a_2 = 1.571 \times 10^2$ and $k_2 = 7.834 \times 10^{-2}$. The fit to the transient region between AD 1820 and 1840 was obtained by polynomial interpolation.

Africa presents an interesting and unique case when the singularity for the hyperbolic trajectory describing the growth of population between AD 1 and 1840 is earlier than the singularity for the hyperbolic trajectory describing the growth of the GDP between AD 1 and 1820. For the growth of the population, the point of singularity is at $t = 2343$ while for the growth of the GDP it is at $t = 2473$ (Nielsen, 2016a, 2016d).

For the linearly-modulated hyperbolic distribution the escape to infinity at a fixed time occurs when the singularity for the growth of the GDP is earlier than the singularity for the growth of population (Nielsen, 2015). If the singularity for the growth of population occurs earlier, as in Africa, then the GDP/cap ratio decreases slowly with time and then escapes rapidly to zero at the time of the singularity for the growth of population. The decreasing GDP/cap distribution between AD 1 and the early 1800s in Africa does not represent an unusual and distinctly different mechanism of economic growth but simply the mathematical property of dividing two hyperbolic distributions describing the growth of the GDP and population. In particular, it does not represent Malthusian stagnation because both the GDP and the population were increasing hyperbolically (Nielsen, 2016a, 2016d).

Africa is also made of less-developed countries (BBC, 2014; Pereira, 2011) so according to Galor (2008a, 2012a) it should have experienced stagnation until around AD 1900 followed by a clear takeoff around that year. These expectations are contradicted by data.

In contradiction of Galor's interpretations of economic growth (Galor, 2005a, 2008a, 2011, 2012a), the Malthusian regime of stagnation did not exist. The GDP and population were increasing hyperbolically during the entire time of the alleged but non-existent regime of Malthusian stagnation, from AD 1 to 1900 and even after that year. Unrecognised by Galor (because he did not analyse data but preferred to use distorted diagrams) there was a transition between two hyperbolic trajectories during his assumed but non-existent regime of Malthusian stagnation reflected in the transition from a slowly decreasing to a fast increasing linearly-modulated hyperbolic trajectory.

Africa is the only region where the economic growth was boosted at the time of the Industrial Revolution but it is also the poorest region, where the claimed Malthusian stagnation should have been most clearly demonstrated. According to Galor, Malthusian stagnation should have prevailed in Africa until around 1900 (Galor, 2008a, 2012a). This hypothesis, which appears to have been confirmed by his manipulation of data, is clearly and convincingly contradicted by their mathematical analysis.

Analysis of data describing the GDP and population in Africa shows that there was no stagnation over the entire range of the AD era (Nielsen, 2016a, 2016d). Economic growth (as described by the GDP) and the growth of population were following the steadily-increasing and undisturbed hyperbolic trajectories but around the time of the Industrial Revolution they were diverted to faster hyperbolic trajectories. There are no signs of Malthusian stagnation before and after the Industrial Revolution and before AD 1900, which was supposed to mark the end of the epoch of Malthusian stagnation. Hyperbolic growth, even if slow, does not represent Malthusian stagnation. Convincing signature of Malthusian stagnation is random fluctuations often described as Malthusian oscillations. This signature is missing in the data but the data show steadily-increasing hyperbolic distributions describing economic growth and the growth of population (Nielsen, 2016a, 2016d).

Analysis of the GDP/cap data shows that after a transition from a slowly-decreasing trajectory before around 1840 (which as we have pointed out does not represent Malthusian stagnation but the mathematical properties of dividing monotonically-increasing hyperbolic trajectories) the growth of the GDP/cap in Africa was following a vigorously-increasing trajectory during the alleged Malthusian stagnation. So while the suitable manipulation of data (Galor, 2005a, 2011) appears to be confirming preconceived ideas, mathematical analysis of precisely the same data shows that the preconceived ideas are clearly incorrect.

This analysis also shows that difficult and primitive living conditions should not be immediately interpreted as Malthusians stagnation. Living conditions in the past may have been primitive and difficult by modern standards but they should not be used as a proof of the existence of Malthusian stagnation. They certainly did not interfere with the economic growth and with the growth of human population (2016a, 2016b, 2016c, 2016d, 2016e, 2016f).

Data and their analyses show no impact and no presence of the hypothetical Malthusian stagnation. While the time-range of the data describing economic growth is relatively short, because it extends only down to AD 1, the time-range of the population data is significantly longer: it extends down to 10,000 BC. Mathematical analysis of data finds no confirmation of the existence of the hypothetical epoch of Malthusian stagnation. It is a vague concept, which has no application to the explanation of the dynamics of economic and demographic growth. Its continuing presence in academic discussions as a tool to explain the dynamics of growth appears to be not only totally irrelevant but also harmful because it diverts attention form finding correct explanations of the growth process.

It is also useful to compare results of the analysis for Africa with the results for Western Europe. Industrial Revolution, "the prime engine of economic growth" (Galor, 2005a, p. 212) should have worked most efficiently in Western Europe and its effects should have been most convincingly confirmed by data, but these effects are convincingly contradicted by data: the alleged engine did not change the economic growth trajectory in Western Europe. Likewise, Malthusian stagnation should have been most prominently confirmed in Africa but is not. There was never any form of stagnation in the economic growth in Africa.

Furthermore, while in Western Europe, Industrial Revolution had absolutely no impact on changing the economic growth trajectory, in Africa there was a spectacular acceleration of growth during the time of the Industrial Revolution but it was not the acceleration from stagnation to growth but from growth to growth. The wished-for features are contradicted by data showing that even plausible stories and explanations should not be accepted in science unless they can be confirmed by relevant data; otherwise they are just stories of fiction.

The alleged sudden acceleration (takeoff) in income per capita is supposed to have been associated with the benefits of progress such as better health care, better housing, better education, higher standard of living and generally better living conditions. However, data show that in Europe there was no takeoff in the income per capita at the time of the Industrial Revolution, while in Africa there was a dramatic acceleration without a dramatic improvement in the style of living. On the contrary, this dramatic boosting in income per capita at the time of the Industrial Revolution appears to coincide with the dramatic deterioration of living conditions of native populations. It occurred around the time of the intensified colonisation of Africa (Duignan & Gunn, 1973; McKay, Hill, Buckler, Ebrey, Beck, Crowston, & Wiesner-Hanks, 2012; Pakenham, 1992).

If a sudden takeoff is supposed to mark the "remarkable" or "stunning" escape from the Malthusian trap (Galor, 2005a, pp. 177, 220), then the only such takeoff occurred in Africa. However, this dramatic takeoff did not mark the transition from stagnation to growth, because there was no stagnation. It also did not mark the dramatic escape from the Malthusian trap because there was no Malthusian trap. (Economic growth and the growth of population were steadily increasing before this takeoff.) It marked the transition from freedom and independence to the trap of misery, deprivation and suffering of the native population of Africa. Story-writing is good in fiction but in science such creative activities should be moderated by the rigorous analysis of relevant data.

## Latin America

Results of analysis of economic growth in Latin America are presented in Figure 8. Data for Latin America contain a discontinuity in the growth of the GDP and population between AD 1500 and 1600 (Nielsen, 2016a, 2016d). This discontinuity is reflected in the discontinuity of the growth of income per capita (GDP/cap).

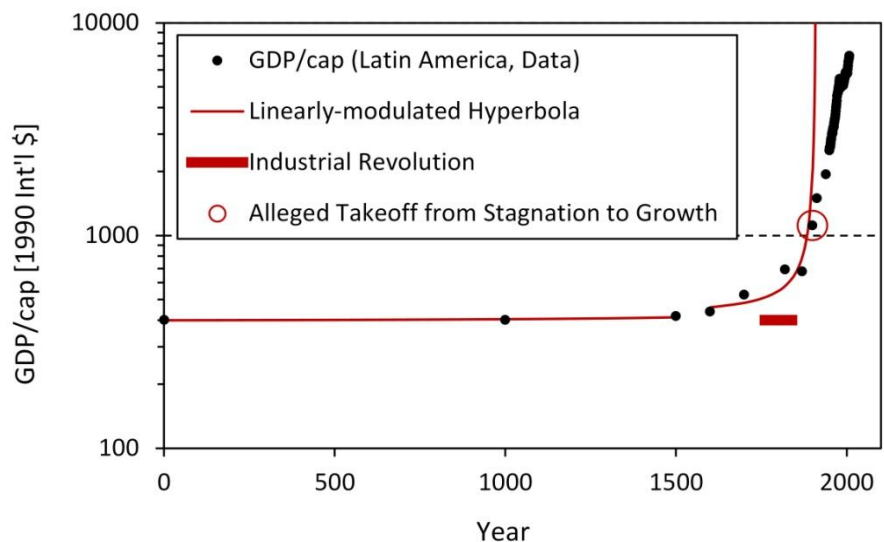

**Figure 8.** *Maddison's data (Maddison, 2010) describing the growth of income per capita (GDP/cap) in Latin America are compared with the linearly-modulated hyperbolic distributions (Nielsen, 2015) obtained by fitting the GDP and population data (Nielsen, 2016a, 2016d). There was a discontinuity in the economic growth and in the growth of population between AD 1500 and 1600 reflected in the discontinuity in the GDP/cap distribution. The alleged takeoff from stagnation to growth around AD 1900 (Galor, 2008a, 2012a) did not happen because the GDP/cap trajectory was not stagnant before that year and because there was no sudden acceleration in growth. On the contrary, around the alleged takeoff the growth of income per capita started to be diverted to a slower trajectory.*

Parameters describing the slowly-increasing linearly-modulated hyperbolic trajectory are $a_1 = 4.421 \times 10^{-1}$ and $k_1 = 2.093 \times 10^{-4}$ for the GDP and $a_2 = 1.765 \times 10^2$ and $k_2 = 8.242 \times 10^{-2}$ for the population. The fast-increasing trajectory from AD 1600 is described by the following parameters: $a_1 = 1.570 \times 10^0$ and $k_1 = 8.224 \times 10^{-4}$ for the GDP and $a_2 = 6.561 \times 10^2$ and $k_2 = 3.371 \times 10^{-1}$ for the population. The discontinuity in the economic growth and in the growth of population coincides with the onset of Spanish conquest (Bethell, 1984). However, after this relatively brief delay, economic growth and the growth of human population were following fast-increasing hyperbolic trajectories.

Latin America is also made of less-developed countries (BBC, 2014; Pereira, 2011) so again, according to Galor (2008a, 2012a), the growth of income per capita (GDP/cap) in this region should have been stagnant until around AD 1900 and fast from around that year. This pattern of growth is contradicted by data. The data show a diametrically different pattern: (1) there is no convincing evidence of the existence of stagnation over the entire range of time between AD 1 and 1900 (convincing evidence of Malthusian stagnation requires the presence of random fluctuations) but there is a sufficiently convincing evidence of the linearly-modulated hyperbolic growth particularly between AD 1600 and 1900; (2) there was no takeoff from stagnation to growth at any time; and (3) at the time of the postulated takeoff in 1900 the growth of income per capita started to be diverted to a slower trajectory. The wished-for takeoff is replaced by a *slower* growth. However, even if we had a takeoff around that time it would have been a takeoff of a different kind, not a takeoff from stagnation to growth as required by the Unified Growth Theory (Galor, 2005a, 2011) but a takeoff from growth to growth.

## Summary and conclusions

Results of mathematical analysis of the historical income per capita (GDP/cap) distributions are presented in Table 1. The listed parameters ($a_1, k_1, a_2, k_2$) describe the fitted, linearly-modulated hyperbolic trajectories (Nielsen, 2015) represented by the ratios of hyperbolic distributions describing the growth of the GDP and population. Parameters $a_1$ and $k_1$ describe hyperbolic distributions fitting the GDP data (Nielsen, 2016a), while parameters $a_2$ and $k_2$ describe hyperbolic distributions fitting population data (Nielsen, 2016d).

**Table 1**. *Summary of the mathematical analysis of the historical income per capita (GDP/cap) distributions*

| Region | $a_1$ | $k_1$ | $a_2$ | $k_2$ | Stagnation | Takeoff |
|---|---|---|---|---|---|---|
| World | $1.684 \times 10^{-2}$ | $8.539 \times 10^{-6}$ | $7.739 \times 10^{0}$ | $3.765 \times 10^{-3}$ | X | X |
| Western Europe | $9.859 \times 10^{-2}$ | $5.112 \times 10^{-5}$ | $7.542 \times 10^{1}$ | $3.749 \times 10^{-2}$ | X | X |
| Eastern Europe | $7.749 \times 10^{-1}$ | $4.048 \times 10^{-4}$ | $3.055 \times 10^{2}$ | $1.525 \times 10^{-1}$ | X | X |
| Former USSR | $6.547 \times 10^{-1}$ | $3.452 \times 10^{-4}$ | $2.618 \times 10^{2}$ | $1.333 \times 10^{-1}$ | X | X |
| Asia | $2.303 \times 10^{-2}$ | $1.129 \times 10^{-5}$ | $1.068 \times 10^{1}$ | $4.999 \times 10^{-3}$ | X | X |
| Africa | $1.244 \times 10^{-1}$ | $5.030 \times 10^{-5}$ | $5.794 \times 10^{1}$ | $2.473 \times 10^{-2}$ | X | X |
| Latin America | $4.192 \times 10^{-1}$ | $2.126 \times 10^{-4}$ | $1.571 \times 10^{2}$ | $7.834 \times 10^{-2}$ | | |
| | $4.421 \times 10^{-1}$ | $2.093 \times 10^{-4}$ | $1.765 \times 10^{2}$ | $8.242 \times 10^{-2}$ | X | X |
| | $1.570 \times 10^{0}$ | $8.224 \times 10^{-4}$ | $6.561 \times 10^{2}$ | $3.371 \times 10^{-1}$ | | |

**Notes:** $a_1, k_1, a_2, k_2$ – Parameters describing linearly-modulated hyperbolic distributions (ratios of hyperbolic distributions). Parameters $a_1$, $k_1$ describe hyperbolic growth of the GDP, while $a_2$, $k_2$ describe hyperbolic growth of population [see eqn (1)]. X – No stagnation/takeoff. Within the range of the mathematically-analysable data the claimed by Galor (2005A, 2008A, 2011, 2012A) Malthusian regime of stagnation did not exist. The claimed takeoffs from stagnation to growth never happened.

Results of this analysis demonstrate explicitly that the postulated by Galor (2005a, 2008a, 2011, 2012a) takeoffs in the income per capita (GDP/cap) did not happen.

There were no transitions from stagnation to growth because within the mathematically-analysable data Galor's regime of Malthusian stagnation did not exist. Growth of income per capita was following the linearly-modulated hyperbolic distributions until recently when it started to be diverted to slower trajectories.

Galor's Unified Growth Theory (Galor, 2005a, 2011) is contradicted yet again by data. The "remarkable" and "stunning' escape from the Malthusian trap (Galor, 2005a, pp. 177, 220) never happened because there was no trap. His claim of the existence of the differential takeoffs is also contradicted by data because we cannot have differential takeoffs without takeoffs. Galor describes phenomena that did not exist. His explanations of economic growth are based on phantom features created by hyperbolic illusions and magnified by his habitually distorted presentation of data such as illustrated in Figure 1. His theory is irrelevant and misleading.

Galor had access to the excellent data of Maddison (2001). He even used them during the formulation of his theory but he did not attempt to analyse them, which is surprising because their analysis is trivially simple (Nielsen, 2014). Now, precisely the same data can be used to demonstrate that his Unified Growth Theory (Galor, 2005a, 2011) is repeatedly contradicted by data (Nielsen, 2014, 2015, 2016a, 2016c, 2016d, 2016e, 2016f).

Unified Growth Theory is fundamentally incorrect and scientifically unacceptable. It is a theory based on scientifically unsupported concepts created by impressions and reinforced by the manipulation of data. Excellent data of Maddison (2001) were not analysed but presented repeatedly using distorted and misleading diagrams such as shown in Figure 1. Such distorted presentation of data appears not only ubiquitously in the Unified Growth Theory but also in other related publications (Ashraf, 2009; Galor, 2005a, 2005b, 2007, 2008a, 2008b, 2008c, 2010, 2011, 2012a, 2012b, 2012c; Galor and Moav, 2002; Snowdon & Galor, 2008). Selected values of data were also repeatedly quoted to support the concept of stagnation followed by takeoffs from stagnation to growth. This is an unscientific approach to research but it is a method used often when defending doctrines accepted by faith.

Galor appears to have been purposefully manipulating evidence to support his preconceived ideas. However, an alternative explanation is that he simply did not know how to analyse data, but this conclusion is hard to accept because he appears to be familiar with mathematics and the analysis of hyperbolic distributions is trivially simple (Nielsen, 2014).

Assisted by the excellent data of Maddison (2001) available to him at the time of the formulation of his theory, Galor was on the verge of making important discovery that economic growth was hyperbolic and thus that there was no Malthusian stagnation and no takeoffs from stagnation to growth. However he missed this first-rate opportunity because he failed to follow the fundamental principles of scientific investigation, which require that theories should be tested by data and that research should be guided and moderated by data.

Here we come to the third and probably the most plausible explanation why Galor appears to have been reluctant to be guided by data and why he apparently manipulated data to support his preconceived ideas. It is what is commonly called being blinded by prejudice or what psychologists describe as cascade behaviour, information cascade, informational avalanche, illusion of truth, illusory truth, illusion of familiarity, running with the pack, following the crowd, herding behaviour, bandwagons and path depending choice (Anderson & Holt, 1997; Begg, Anas & Farinacci, 1992; Bikhchandani, Hirshleifer & Welch, 1992, 1998; De Vany & Lee, 2008; De Vany & Walls, 1999; Easley & Kleinberg, 2010; Grebe, Schmid & Stiehler, 2008; Ondrias, 1999; Parks & Tooth, 2006; Ramsey, Raafat, Chater & Frith, 2009; Walden & Browne, 2003).

In the demographic and economic research this phenomenon is demonstrated by the reluctance to accept the compelling contradicting evidence in data simply because many demographers or economists would not agree with the contradicting evidence. It is safer to follow the crowd and run with the pack. Tradition is stronger than science and only an outsider who has not been blinded by prejudice and who is not afraid of being rejected by the crowd might dare to show that the accepted doctrines are incorrect. He or she is then likely to be ridiculed and rejected but science is a self-correcting discipline so sooner or later such resistance to accept the overwhelming empirical evidence will have to be broken, but it would be better for science and scientists if the required change in the paradigm is accepted sooner rather than later.

The evidence is overwhelming: historical economic growth and historical growth of population were hyperbolic (Kapitza, 2006; Kremer, 1993; Nielsen, 2014, 2015, 2016a, 2016b, 2016c, 2016d, 2016e, 2016f; Podlazov, 2002; Shklovskii, 1962, 2002; von Foerster, Mora and Amiot, 1960; von Hoerner, 1975). Hyperbolic growth should be the basis for explaining the mechanism of the historical growth of population and of the historical economic growth.

Interpretations revolving around the concept of Malthusian stagnation and around transitions from stagnation to growth are repeatedly and consistently contradicted by data and by their mathematical analyses. Historical economic growth and historical growth of population cannot be divided into distinctly different regimes governed by distinctly different mechanisms of growth. Hyperbolic growth has to be explained as a whole. The same mechanism has to be applied to the perceived slow and fast components because it is mathematically impossible to divide hyperbolic distributions into distinctly different sections (Nielsen, 2014). Once we can explain properly the mechanism of the past growth we might be able to understand better the current growth and how it should be controlled.

# References

Anderson, L. R., & Holt, C. A. (1997). Information cascades in the laboratory. *American Economic Review, 87*(5), 847-862.

Ashraf, Q. H. (2009). Essays on Deep Determinants of Comparative Economic Development. Ph.D. Thesis, Department of Economics, Brown University, Providence.

BBC, (2014). The North South Divide. http://www.bbc.co.uk/bitesize/standard/geography/international_issues/contrasts_development/revision/2/

Begg, I. M., Anas, A., & Farinacci, S. (1992). Dissociation of process in belief: Source recollection, statement familiarity and the illusion of truth. *Journal of Experimental Psychology. 121*(4), 446 – 458.

Bethell, L. (Ed.). (1984). *The Cambridge History of Latin America*: Vol. I and II, Colonial Latin America. Cambridge, UK: Cambridge University Press.

Biraben, J-N. (1980). An Essay Concerning Mankind's Evolution. Population, Selected Papers, December.

Bikhchandani, S., Hirshleifer, D., & Welch, I, A. (1992). Theory of fads, fashion, custom, and cultural change as informational cascades. *Journal of Political Economy, 100*(5), 992-1026.

Clark, C. (1968). *Population Growth and Land Use*. New York, NY: St Martin's Press.

Cook, R. C. (1960). World Population Growth. *Law and Contemporary Problems, 25*(3), 379-388.

De Vany, A., & Lee, C. (2008). Information cascades in multi-agent models. http://pages.stern.nyu.edu/~wgreene/entertainmentandmedia/Cascades.pdf

Duignan, P., & Gunn, L. H. (Eds.) (1973). *Colonialism in Africa 1870 – 1960: A Bibliographic Guide to Colonialism in Sub-Saharan Africa*. Cambridge, UK: Cambridge University Press.

Durand, J. (1967). A Long-range View of World Population Growth. *The Annals of the American Academy of Political and Social Science: World Population, 369*, 1-8.

Durand, J. D. (1974). Historical Estimates of World Population: An Evaluation. Analytical and Technical Reports, Number 10. University of Pennsylvania, Population Center.

Durand, J. (1977). Historical Estimates of World Population: An Evaluation. *Population and Development Review, 3*(3), 256-293.

Floud, D., & McCloskey, D. N. (1994). *The Economic History of Britain since 1700*. Cambridge: Cambridge University Press.

Easley, D., & Kleinberg, J. (2010). *Networks, crowds, and markets*. Cambridge, UK: Cambridge University Press.

Gallant, R. A. (1990). *The Peopling of Planet Earth: Human Growth through the Ages*. New York, NY: Macmillan Publishing Company.

Galor, O. (2005a). From stagnation to growth: Unified Growth Theory. In P. Aghion & S. Durlauf (Eds.), *Handbook of Economic Growth* (pp. 171-293). Amsterdam: Elsevier.

Galor, O. (2005b). The Demographic Transition and the Emergence of Sustained Economic Growth. *Journal of the European Economic Association, 3*, 494-504. http://dx.doi.org/10.1162/jeea.2005.3.2-3.494

Galor, O. (2008a). Comparative Economic Development: Insight from Unified Growth Theory. http://www.econ.brown.edu/faculty/Oded_Galor/pdf/Klien%20lecture.pdf

Galor, O. (2008b). Economic Growth in the Very Long Run. In: Durlauf, S.N. and Blume, L.E., Eds., *The New Palgrave Dictionary of Economics*. Palgrave Macmillan, New York. http://dx.doi.org/10.1057/9780230226203.0434

Galor, O. (2008c). Comparative Economic Development: Insight from Unified Growth Theory. http://www.econ.brown.edu/faculty/Oded_Galor/pdf/Klien%20lecture.pdf

Galor, O. (2010). The 2008 Lawrence R. Klein Lecture—Comparative Economic Development: Insights from Unified Growth Theory. International Economic Review, 51, 1-44. http://dx.doi.org/10.1111/j.1468-2354.2009.00569.x

Galor, O. (2011). *Unified Growth Theory*. Princeton, New Jersey: Princeton University Press.

Galor, O. (2012a). Unified Growth Theory and Comparative Economic Development. http://www.biu.ac.il/soc/ec/students/mini_courses/6_12/data/UGT-Luxembourg.pdf

Galor, O. (2012b). The Demographic Transition: Causes and Consequences. *Cliometrica, 6*, 1-28. http://dx.doi.org/10.1007/s11698-011-0062-7

Galor, O. (2012c). Unified Growth Theory and Comparative Economic Development. http://www.biu.ac.il/soc/ec/students/mini_courses/6_12/data/UGT-Luxembourg.pdf

Galor, O. and Moav, O. (2002). Natural Selection and the Origin of Economic Growth. *The Quarterly Journal of Economics, 117*, 1133-1191. http://dx.doi.org/10.1162/003355302320935007

Grebe, T., Schmid, J., & Stiehler, A. (2008). Do individuals recognize cascade behavior of others? – An experimental study. *Journal of Economic Psychology, 29*(2), 197–209.

Haub, C. (1995). How Many People Have Ever Lived on Earth? *Population Today*, February.

Kapitza, S. P. (2006). *Global population blow-up and after*. Hamburg: Global Marshall Plan Initiative.

Kremer, M. (1993). Population Growth and Technological Change: One Million B.C. to 1990. *Quarterly Journal of Economics, 108*(3), 681–716.

Livi-Bacci, M. (1997). *A Concise History of World Population* (2nd ed.). Malden, MA: Blackwell Publishers.

Maddison, A. (2001). *The World Economy: A Millennial Perspective*. Paris: OECD.

Maddison, A. (2010). Historical Statistics of the World Economy: 1-2008 AD. http://www.ggdc.net/maddison/Historical Statistics/horizontal-file_02-2010.xls.

McEvedy, C., & Jones, R. (1978). *Atlas of World Population History*. Middlesex, England: Penguin.

Nielsen, R. W. (2014). Changing the Paradigm. *Applied Mathematics, 5*, 1950-1963. http://dx.doi.org/10.4236/am.2014.513188

Nielsen, R. W. (2015). Unified Growth Theory contradicted by the GDP/cap data. http://arxiv.org/ftp/arxiv/papers/1511/1511.09323.pdf

Nielsen, R. W. (2016a). Mathematical analysis of the historical economic growth with a search for takeoffs from stagnation to growth. *Journal of Economic Library, 3*(1), 1-23. http://www.kspjournals.org/index.php/JEL/article/view/606

Nielsen, R. W. (2016b). Growth of the world population in the past 12,000 years and its link to the economic growth. *Journal of Economics Bibliography, 3*(1), 1-12. http://www.kspjournals.org/index.php/JEB/article/view/607

Nielsen, R. W. (2016c). The postulate of the three regimes of economic growth contradicted by data. *Journal of Economic and Social Thought, 3*(1), 1-34. http://www.kspjournals.org/index.php/JEST/article/view/669

Nielsen, R. W. (2016d). Unified Growth Theory contradicted by the mathematical analysis of the historical growth of human population. http://arxiv.org/ftp/arxiv/papers/1601/1601.07291.pdf. *Journal of Economics and Political Economy – forthcoming.*

Nielsen, R. W. (2016e). Unified Growth Theory contradicted by the absence of takeoffs in the Gross Domestic Product. *Economic Review, 3*(1), 16-27. http://www.kspjournals.org/index.php/TER/article/view/650

Nielsen, R. W. (2016f). Puzzling properties of the historical growth rate of income per capita explained. http://arxiv.org/ftp/arxiv/papers/1603/1603.00736.pdf. *Journal of Economics Library - forthcoming*

McKay, J. P., Hill, B. D., Buckler, J., Ebrey, P. B., Beck, R. B., Crowston, C. H., & Wiesner-Hanks, M. E. (2012). *A History of World Societies: From 1775 to Present*. Volume C – From 1775 to the Present. Ninth edition. Boston, MA: Bedford Books.

Pakenham, T. (1992). *The Scramble for Africa: White Man's Conquest of the Dark Continent from 1876-1912*. New York: Avon Books.

Ondrias, K. (1999). *The brain, consciousness & illusion of truth* (E. Nazinska, Trans. & Ed.). Boca Raton, FL: Universal Publishers.

Parks, C. M., & Tooth, J. P. (2006). Fluency, familiarity, aging, and the illusion of truth. *Aging, neuropsychology, and cognition, 13*, 225–253

Pereira, E. (2011). Developing Countries Will Lead Global Growth in 2011, Says World Bank. http://www.forbes.com/sites/evapereira/2011/01/12/developing-countries-will-lead-global-growth-in-2011-says-world-bank/

Podlazov, A.V. (2002). Theoretical demography: Models of population growth and global demographic transition (in Russian). In *Advances in Synergetics: The Outlook for the Third Millennium* (pp. 324–345). Moscow: Nauka.

Shklovskii, J. S. (1962). *The universe, life and mind*, (in Russian). Moscow: Academy of Science, USSR.

Shklovskii, J. S. (2002*). The universe life and mind* (5th edn.). John Wiley and Sons, Ltd, New York, US.

Ramsey M., Raafat, R. M., Chater, N., & Frith, C. (2009). Herding in humans. *Trends in Cognitive Sciences, 13*(10), 420-428.

Snowdon, B., & Galor, O. (2008). Towards a Unified Theory of Economic Growth. *World Economics, 9*, 97-151.

Taeuber, C., & Taeuber, I. B. (1949). World Population Trends. *Journal of Farm Economics, 31*(1), 241.

Thomlinson, R. (1975). *Demographic Problems, Controversy Over Population Control* (2nd ed.). Encino, Ca.: Dickensen Pub.

Trager, J. (1994). *The People's Chronology: A Year-by-Year Record of Human Events from Prehistory to the Present*. New York, NY: Henry Holt and Company.

von Foerster, H., Mora, P., & Amiot, L. (1960). Doomsday: Friday, 13 November, A.D. 2026. *Science, 132*, 1291-1295.

von Hoerner, S. J. (1975). Population explosion and interstellar expansion. *Journal of the British Interplanetary Society, 28*, 691-712.

Walden, E. A., & Browne, G. J. (2003). *Running with the pack: An empirical examination of information cascades in the adoption of novel technologies*. Lubbock: Texas Tech. University.